\documentclass{ptephy}

\begin{document}

\title{Mass and radius formulas for low-mass neutron stars}

\author{\name{Hajime Sotani}{1,\ast}, \name{Kei Iida}{2}, \name{Kazuhiro Oyamatsu}{3}, and \name{Akira Ohnishi}{1}}

\address{\affil{1}{Yukawa Institute for Theoretical Physics, Kyoto University, Kyoto 606-8502, Japan}
\affil{2}{Department of Natural Science, Kochi University, 2-5-1 Akebono-cho, Kochi 780-8520, Japan}
\affil{3}{Department of Human Informatics, Aichi Shukutoku University, 9 Katahira, Nagakute, Aichi 480-1197, Japan}
\email{sotani@yukawa.kyoto-u.ac.jp}}

\begin{abstract}
Neutron stars, produced at the death of massive stars, are often
regarded as giant neutron-rich nuclei.  This picture is especially 
relevant for low-mass (below about solar mass, $M_\odot$)
neutron stars, where non-nucleonic components are not expected
to occur.  Due to the saturation property of nucleonic matter, 
leading to the celebrated liquid-drop picture of atomic nuclei, empirical 
nuclear masses and radii can be approximately expressed as 
function of atomic mass number.  It is, however, not 
straightforward to express masses and radii of neutron stars even in the 
low-mass range where the structure is determined by a balance between the 
pressure of neutron-rich nucleonic matter and the gravity.  Such 
expressions would be of great use given possible simultaneous mass and radius 
measurements.  Here we successfully construct theoretical formulas 
for the masses and radii of low-mass neutron stars from various 
models that are consistent with empirical masses and radii of stable nuclei. 
In this process, we discover a new equation-of-state
parameter that characterizes the structure of low-mass neutron stars.  
This parameter, which plays a key role in connecting the
mass-radius relation of the laboratory nuclei to that of the celestial objects,
could be constrained from future observations of low-mass neutron stars.
\end{abstract}

\subjectindex{E32, D41}

\maketitle

\section{Introduction}

Neutron stars have been serving as laboratories to probe the densest 
and most neutron-rich matter in the Universe.
It is generally believed that the outer, low-density part 
of a neutron star (crust) consists of a body-center-cubic lattice 
of neutron-rich nuclei, embedded in a gas of electrons and, if any,  
dripped neutrons, and near normal nuclear density ($\rho_0$), the
nuclei melt into uniform nucleonic matter, which mainly 
composes the star's core \citep{LaPr}.  The equation of state (EOS) of matter in the star, i.e.,
neutron star matter, has one-to-one correspondence to the star's
mass ($M$) and radius ($R$) relation via hydrostatic 
equilibrium.  Observational data for $M$ have been 
accumulated \citep{LaPr,KKYT}, whereas those for $R$ have been recently 
estimated from observations of thermonuclear X-ray bursts 
with photospheric radius expansion and thermal spectra from
quiescent low-mass X-ray binaries
\citep{SLB2012,OBG2010,GSWR2013,LS2013}.

In theoretically describing laboratory nuclei and
neutron star matter, it is useful to consider the energy of 
``nuclear matter,'' i.e., hypothetical infinite matter, composed 
of neutrons and of protons that have electric charge
switched off.  For simplicity, as neutron star matter, 
we will consider zero-temperature, $\beta$ equilibrated,
charge neutral matter made of real nucleons and electrons.
The EOS of nuclear matter is still uncertain even 
near $\rho_0$, while it can be constrained from terrestrial 
nuclear experiments \citep{exp} and neutron star 
observations \citep{OBG2010,LS2013,D2010} via theoretical 
calculations.  It is noteworthy that the candidates for
low-mass neutron stars have been discovered in binary 
systems \citep{LP2011}, which could give additional 
information on the EOS once $R$ is measured.

The energy of uniform nuclear matter can be expanded
around the saturation point of symmetric nuclear matter
(SNM), i.e., nuclear matter made of the same number of 
neutrons and protons, with respect to the nucleon number 
density, $n_b$, and neutron excess, $\alpha$,
defined as $\alpha \equiv (n_n-n_p)/n_b$, where $n_n$ and $n_p$ 
denote the neutron and proton number densities. In practice,
in the vicinity of the saturation point of SNM at zero temperature, the  
energy per nucleon, $w$, of uniform nuclear matter can be 
written as a function of $n_b$ and $\alpha$ \citep{L1981}, i.e.,
\begin{equation}
w = w_0  + \frac{K_0}{18n_0^2}(n_b-n_0)^2 + \left[S_0 
         + \frac{L}{3n_0}(n_b-n_0)\right]\alpha^2, \label{eq:w}
\end{equation}
where $w_0$, $n_0$, and $K_0$ are the saturation energy, 
the saturation density, and the incompressibility of SNM, while 
$S_0$ and $L$ are associated with the symmetry energy coefficient $S(n_b)$.
That is, $S_0=S(n_0)$ is the symmetry energy coefficient at $n_b=n_0$, while 
$L$ characterizes the density dependence of the nuclear symmetry energy 
around $n_b=n_0$, defined as $L=3n_0(dS/dn_b)_{n_b=n_0}$.  Among these five 
parameters in Eq. (\ref{eq:w}), $w_0$, $n_0$, and $S_0$
can be relatively easier to determine from empirical data for masses and 
radii of stable nuclei, while the remaining two parameters, $K_0$ and $L$, 
are more difficult to fix \citep{OI2003}.  This is why we focus on the 
various sets of $K_0$ and $L$ (Table \ref{tab:EOS}) 
in analyzing neutron star matter.

\begin {table}
\caption{Nuclear matter EOS parameters}
\label{tab:EOS}
\begin {center}
\begin{tabular}{ccccc}
\hline\hline
EOS
 & $K_0$ (MeV) & $L$ (MeV) & $\eta$ (MeV) &  \\
\hline
OI-EOSs
 &   180 & 31.0   & 55.7    &  \\
 &   180 & 52.2   & 78.9    &  \\
 &   230 & 42.6   & 74.7    &  \\
 &   230 & 73.4   & 107  &  \\
 &   280 & 54.9   & 94.5    &  \\
 &   280 & 97.5   & 139  &  \\
 &   360 & 76.4   & 128  &  \\
 &   360 & 146 & 197 &  \\
\hline
Shen
 & 281 & 114  & 154  & \\
Miyatsu
 & 274 &  77.1 & 118  & \\
\hline
FPS
 & 261 & 34.9 & 68.2  & \\ 
SLy4
 & 230 & 45.9 &  78.5 & \\
BSk19
 & 237 & 31.9 & 62.3  & \\
BSk20
 & 241 & 37.4 & 69.6  & \\
BSk21
 & 246 & 46.6 & 81.1  & \\
\hline
\end{tabular}
\end {center}
\end{table}

These two parameters, $K_0$ and $L$, mainly determine the stiffness of 
neutron-rich nuclear matter, but have yet to be fixed. It is also suggested that
$K_0$ is related to the giant resonances of stable 
nuclei \cite{Blaizot1980}, while $L$ is associated with the 
structure and reactions of neutron-rich 
nuclei \cite{exp,Roca-Maza,OI2003} and the pressure 
of pure neutron matter at the saturation density of 
SNM.  Additionally, one could constrain $L$ via 
quasi-periodic oscillations in giant flares observed 
from soft-gamma repeaters \cite{SNIOa,SNIOb}.

In contrast to the well-known empirical nuclear mass and 
radius formulas \citep{BW}, the neutron star counterparts have 
to be theoretically given as function of not only the central 
density ($\rho_c$), but
such EOS parameters as $K_0$ and $L$.  
So far, however, the dependence of low-mass neutron star models 
on $K_0$ and $L$ remains to be examined systematically.  We 
thus start with construction of the neutron star models 
from various EOSs of neutron star matter that meet the 
following conditions:
\begin{enumerate}
 \item Unified description of matter in the crust 
and core based on the same EOS of nuclear matter with specific 
values of $K_0$ and $L$.  
 \item Consistency of the masses and radii 
of stable nuclei calculated within the same theoretical framework 
with the empirical values. 
\end{enumerate}
Mass and radius formulas for low-mass neutron stars are finally obtained in
such a way as to approximately reproduce the neutron star models thus
constructed.

\section{Adopted EOS's of neutron star matter}

Among many available EOSs of neutron star matter, we adopt 
the EOSs that meet the above conditions, i.e., unified EOSs,
which are categorized into 
three groups as in Table \ref{tab:EOS}.  
The first is based on the phenomenological EOS of uniform nuclear matter 
that was constructed by two of us \citep{OI2003},
using a simplified version of the extended Thomas-Fermi theory \citep{O1993}, 
in such a way as to reproduce empirical masses and radii of stable 
nuclei.   They adopted the Pad{\'e}-type potential energies with 
respect to the nucleon density $n_b$ for SNM and for neutron matter, 
respectively, and connected them in a quadratic approximation with respect 
to neutron excess $\alpha$.  This form of the potential energy can well 
reproduce the variational calculations of 
\citet{FP}, to which, in fact, the high-density 
behavior of neutron matter was adjusted.  The $\alpha$ dependence of the 
potential energy is partially justified by the variational 
calculations of \citet{LP}, and the expression
for the total energy reproduces Eq. (\ref{eq:w}) in the limit of 
$n_b\to n_0$ and $\alpha\to 0$.  With such EOSs of uniform nuclear 
matter obtained for various sets of ($K_0$, $L$), they constructed 
the EOSs of neutron star matter \citep{OI2007} by generalizing
the above Thomas-Fermi 
theory as done by \citet{O1993}.
Hereafter, such EOSs of neutron star matter are referred to as the OI-EOSs.
We remark that generally accepted values of $K_0$ lie in the range of
$230 \pm 40$ MeV \cite{KM2013} or so, while the OI-EOSs include rather
extreme cases of $K_0 = 180$ and 360 MeV, as shown in Table \ref{tab:EOS},
to cover the large parameter space. The final mass and radius formulas would
remain almost unchanged even if the OI-EOSs with $K_0 = 180$ and 360 MeV
are not included in the fitting.

In the second group, there are two EOSs of neutron star matter 
calculated within the relativistic framework.  One is the Shen EOS
based on the relativistic mean field theory with the TM1 nuclear 
interaction \citep{ShenEOSa}, and the other is the Miyatsu EOS
based on the relativistic Hartree-Fock theory with the chiral 
quark-meson coupling model \citep{Miyatsu}.  In both EOSs, the same 
type of the Thomas-Fermi model as used for the OI-EOSs is
used in describing neutron star matter in such a way as to reproduce
empirical masses and radii of stable nuclei.

The third group is composed of the five EOSs of neutron star matter 
based on the Skyrme-type effective interactions: FPS \citep{FPS}, 
SLy4 \citep{SLy4}, BSk19, BSk20, and
BSk21 \citep{BSk,PGC2011,PCGD2012}. The FPS
interaction, which was constructed by fitting the properties of 
uniform nucleon matter calculated by \citet{FP}, 
well reproduces the empirical ground-state properties of doubly magic
stable nuclei via the Hartree-Fock calculations.  The SLy4 interaction 
was constructed by \citet{Chabanat1997}
in such a way as to reproduce the microscopic EOS of 
neutron matter calculated with the UV14+UVII nuclear force by \citet{WFF}
as well as the empirical ground-state properties of 
doubly magic stable nuclei within the Hartree-Fock 
approximation.  The BSk19, BSk20, and BSk21
interactions are written in the form of the nuclear energy-density 
functionals, which are derived from generalized Skyrme interactions 
in such a way as to fit all the available nuclear mass data \citep{BSk}.
As a result,
empirical charge radii were also well reproduced.   These interactions
are different in the sense that BSk19, BSk20, and BSk21 are fitted to 
the EOSs of neutron matter derived by \citet{FP},
\citet{Akmal}, and \citet{LS}, respectively.
This difference is expected to play a role in estimating the effect of
uncertainties in three-neutron interactions on the stellar properties, as we shall see.
In describing neutron star matter, a compressible liquid-drop approach was used
for FPS and SLy4, while an extended Thomas-Fermi model was used for BSk19, BSk20
and BSk21.  To calculate the neutron star models in the present study, 
we adopt the analytical expressions for FPS and SLy4 given by 
\citet{HP2004} and for BSk19, BSk20, and BSk21 given by \citet{PFCPG}.

\begin{figure}
  \centering
  \includegraphics[width=10.0cm]{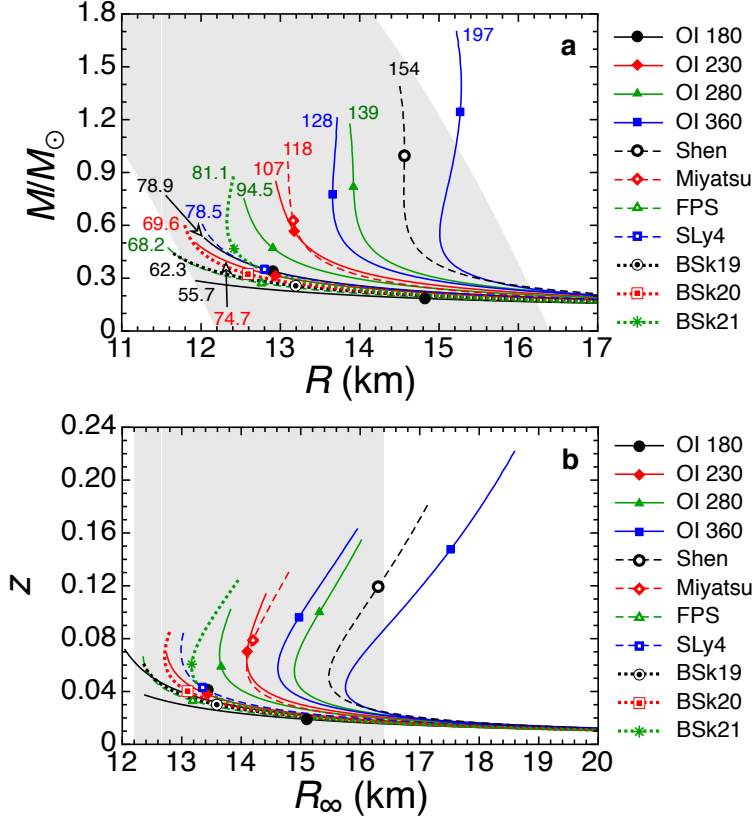}
  \caption{Neutron star properties.
The stellar models are constructed from various unified EOSs
with different sets of $(L,K_0)$.  We plot the relations between the mass 
and radius (a) and between the gravitational redshift and 
radiation radius (b).  The mark and end on each line denote
the stellar models with $\rho_c=1.5\rho_0$ and $2.0\rho_0$,
respectively.
In (a), the labels on the lines denote the values
of the nuclear matter parameter $\eta$.
To distinguish between the OI-EOSs,
we add the values of $K_0$ to the OI-EOS labels; for example, we use 
``OI 180" for the two OI-EOSs with $K_0=180$ MeV (left, smaller $L$; right, 
larger $L$). The shaded region corresponds to the allowed region from the
observed radiation radius of the neutron star in $\omega$ Cen (see text for 
details). }
\label{fig:mr}
\end{figure}

\section{Neutron star models}

Now, we construct nonrotating neutron stars by
integrating the Tolman-Oppenheimer-Volkoff equations from the stellar 
center of density $\rho_c$ outward up to the position where 
the pressure vanishes.  It is not clear up to what density 
the adopted unified EOSs are applicable.  Nonetheless, one can expect
that non-nucleonic components such as hyperons and 
quarks do not occur below $\sim2\rho_0$ \citep{LaPr} and
that the uncertainty from three-neutron interactions in the EOS of 
pure neutron matter becomes relevant above $\sim2\rho_0$,
as suggested by quantum Monte Carlo (QMC) calculations \citep{GCR2012}.
We thus examine the stellar models for $\rho_c\le 2\rho_0$,
where $\rho_0$ is set to $2.68\times 10^{14}$ g cm$^{-3}$, 
and the resultant $M$-$R$ relations are plotted in 
Fig.\ \ref{fig:mr}(a).

To systematically describe various stellar models, 
we introduce a new auxiliary parameter $\eta$ defined as 
$\eta=(K_0L^2)^{1/3}$. 
The values of $\eta$ are shown in Table \ref{tab:EOS}.
Remarkably, the $M$-$R$ relation changes almost smoothly with $\eta$.
Note that the OI-EOSs \citep{OI2007} with $L\lesssim 10$ MeV
are too soft to keep the pressure positive and thus not used here.
This implies the lower limit of $\eta$ of order 30 MeV.
Meanwhile, the EOS models used here cover the values of $\eta$ up to 
$\sim200$ MeV, which is significantly larger than expected from 
existing nuclear experiments.
We remark that the powers of $L$ and $K_0$ in $\eta$ are chosen
to be simple rational numbers in such a way that $\eta$ has the same 
unit as $L$ and $K_0$, i.e., MeV. If one considers arbitrary real numbers
as the exponents, therefore, one could choose different kinds of $\eta$
with which the $M$-$R$ relation changes as smoothly as the present choice.

From the observational viewpoint, the radiation radius 
$R_\infty = R/\sqrt{1-2GM/Rc^2}$ and the gravitational redshift 
$z = 1/\sqrt{1-2GM/Rc^2}-1$ with the gravitational constant $G$ 
and the speed of light $c$ could be more relevant in describing the 
stellar properties than $M$ and $R$.  The calculated $z$-$R_\infty$ 
relation again shows a smooth change with $\eta$ (Fig.\ \ref{fig:mr}(b)).
The photon flux, if detected, would be proportional to 
$(R_\infty/D)^2$, where $D$ is the distance from the Earth, while
the gravitational redshift could be determined from the 
possible shift of atomic absorption lines in spectra of 
the stars.

The smooth change of the stellar properties with $\eta$ suggests that 
not only future nuclear experiments but also simultaneous 
measurements of $M$ and $R$ or, equivalently, $z$ and $R_\infty$
could constrain $\eta$, which could in turn lead to restriction of the
stellar models.  In particular, observations of low-mass neutron stars 
would be essential.  For example, the radiation radius of the X-ray source, 
CXOU 132619.7--472910.8, in the globular cluster NGC 5139 ($\omega$ Cen) 
has been determined as $R_\infty = 14.3\pm 2.1$ km from the $Chandra$ 
data \citep{RBBPZ2002}.  The allowed region from this $R_\infty$ is
shown in Fig.\ \ref{fig:mr}(a) and (b) with the shaded region.  This is 
still consistent with various values of $\eta$, but future precise
determination of $R_\infty$ could constrain $\eta$, if $M$ is low enough. 
Additionally, thermal spectra detected from quiescent low-mass X-ray binaries 
are expected to give $M$ and $R$ simultaneously \citep{GSWR2013,LS2013}.

\begin{figure}
  \centering
  \includegraphics[width=10.0cm]{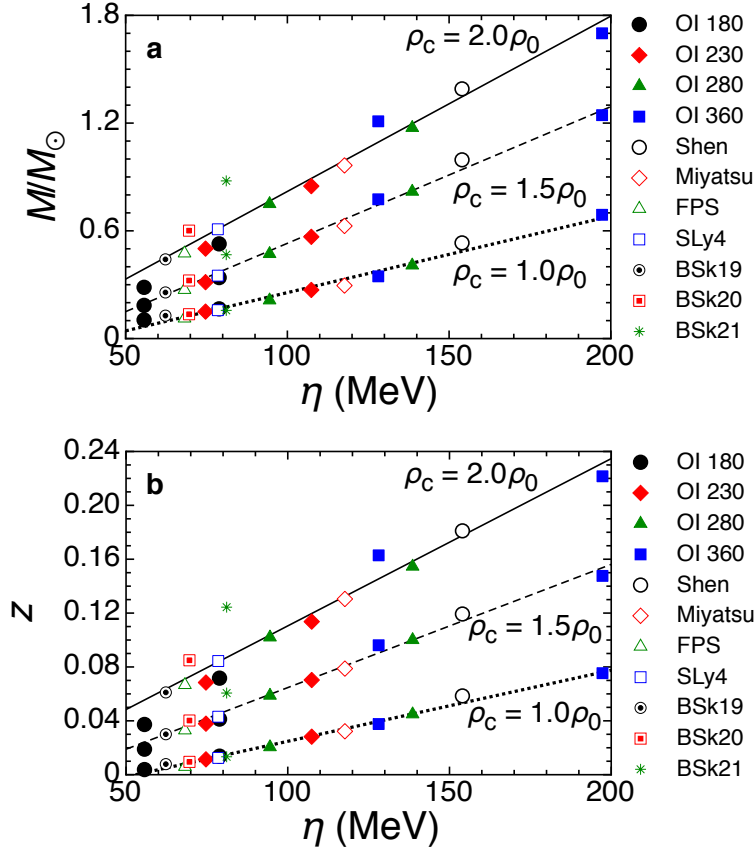}
  \caption{Neutron star masses (a) and
gravitational redshifts (b) as a function of $\eta$.
The stellar models constructed from various unified EOSs are 
given for $\rho_c=2.0\rho_0$, $1.5\rho_0$, and $1.0\rho_0$.
The solid, broken, and dotted lines are the linear fitting to 
the cases of $\rho_c=2.0\rho_0$, $1.5\rho_0$, and $1.0\rho_0$,
respectively (see text for details).}
\label{fig:etaM}
\end{figure}

\section{Mass and radius formulas}

\begin{figure}
  \centering
  \includegraphics[width=10.0cm]{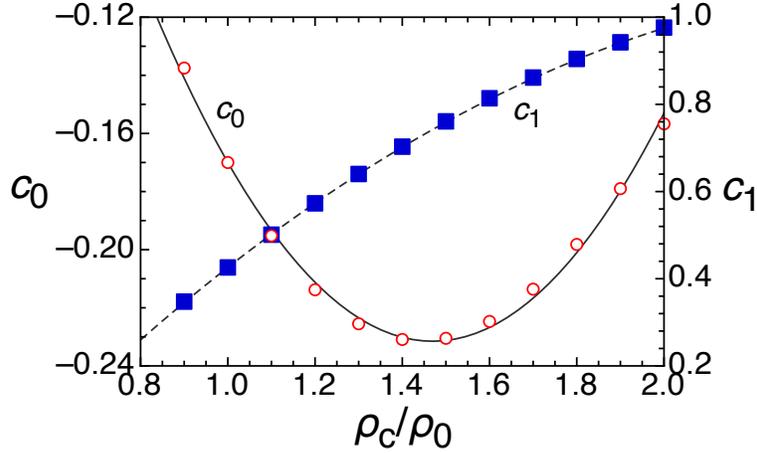}
  \caption{Values (marks) of the adjustable parameters $c_0$ and $c_1$ 
in the mass formula.  The corresponding quadratic fitting curves 
(solid and broken lines) are also shown as a function of $\rho_c/\rho_0$.  
Here we consider the stellar models only for $\rho_c\gtrsim 0.9\rho_0$ 
to avoid unstable neutron star models.}
\label{fig:c0c1}
\end{figure}

To examine the dependence of the stellar properties on $\eta$
more clearly, we plot the stellar masses calculated for
$\rho_c=2.0\rho_0$, $1.5\rho_0$, and $1.0\rho_0$ 
(Fig.\ \ref{fig:etaM}(a)).  From this figure, we find that the 
stellar masses for fixed $\rho_c$ can be approximately expressed 
as a linear function of $\eta$,
$M/M_\odot =   c_0+ c_1(\eta/100\, {\rm MeV})$, 
where $c_0$ and $c_1$ are adjustable parameters that depend on 
$\rho_c$.  The validity of $\eta$ is now evident.  The deviation 
of the calculations from the linear fit at 
$\rho_c=2.0\rho_0$ is larger than that at $\rho_c=1.0\rho_0$,  
particularly for BSk20 and BSk21.  Such deviation is of the 
order of uncertainties in $M$ due to three-neutron interactions obtained
from the QMC evaluations \citep{GCR2012}.
The parameters $c_0$ and $c_1$ can then be expressed
as a quadratic function of $u_c \equiv\rho_c/\rho_0$ within the 
accuracy of errors less than a few percent (Fig.\ \ref{fig:c0c1}).
Finally, we obtain the mass formula:
\begin{equation}
  \frac{M}{M_\odot} =   0.371 - 0.820 u_c + 0.279 u_c^2
     - (0.593 - 1.25 u_c + 0.235 u_c^2) \left(\frac{\eta}{100\, {\rm MeV}}\right), \label{eq:Mass}
\end{equation}
where we confine ourselves to $\rho_c \gtrsim 0.9\rho_0$; otherwise,
the stellar models can become unstable with respect to decompression,
depending on the EOS of neutron star matter.

We also find that the gravitational redshift calculated for fixed 
$\rho_c$ can be approximately expressed as a linear function of $\eta$
(Fig.\ \ref{fig:etaM}(b)).  Then, just like the mass formula
(\ref{eq:Mass}), we can obtain the theoretical formula for $z$ as
\begin{equation}
  z =   0.00859 - 0.0619 u_c + 0.0255 u_c^2
     - (0.0429 - 0.108 u_c + 0.0120 u_c^2) \left(\frac{\eta}{100\, {\rm MeV}}\right). \label{eq:zz}
\end{equation}
Using Eqs. (\ref{eq:Mass}) and (\ref{eq:zz}), one could estimate the values
of $\eta$ and $u_c$ from possible simultaneous measurements of $M$ and $z$.
In general, Eqs. (\ref{eq:Mass}) and (\ref{eq:zz}) can have as many as four sets of
solutions $(u_c, \eta)$ for given observational values of $M/M_\odot$ and $z$.
As mentioned above, however, Eqs. (\ref{eq:Mass}) and (\ref{eq:zz}) are valid in
the range of $0.9 \lesssim u_c \le 2.0$.  In this range, as shown in Fig. \ref{fig:etaM},
the solution $(u_c, \eta)$ has to be unique.

\begin{figure}
  \centering
  \includegraphics[width=10.0cm]{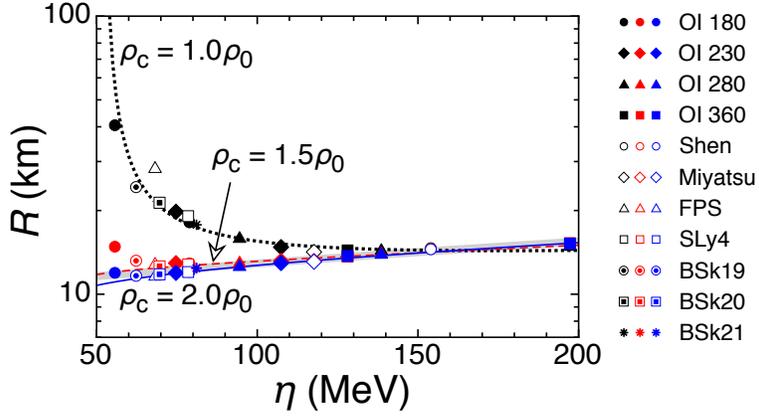}
  \caption{Neutron star radii as a function of $\eta$.
The stellar models constructed from various unified EOSs are 
given for $\rho_c=1.0\rho_0$ (black), $1.5\rho_0$ (red),
and $2.0\rho_0$ (blue). 
The solid, broken, and dotted lines are the formula values for
the cases of $\rho_c = 2.0\rho_0$, 1.5$\rho_0$, and 1.0$\rho_0$,
respectively, obtained from Eqs. (\ref{eq:Mass}) and (\ref{eq:zz}).
The thick straight line denotes the converging behavior expressed by 
Eq. (\ref{eq:fit-R}).}
\label{fig:etaR}
\end{figure}

It is straightforward to obtain the formula for $R$ from Eqs. 
(\ref{eq:Mass}) and (\ref{eq:zz}).  The obtained formula can be compared
with the calculations of $R$ for $\rho_c=1.0\rho_0$, $1.5\rho_0$, and 
$2.0\rho_0$ (Fig.\ \ref{fig:etaR}).  
We confirm a good agreement 
between those two except for $\eta\lesssim70$ MeV.  The mass and radius 
formulas could help to constrain not only the nuclear matter parameter 
$\eta$ but also a star's $\rho_c$ via possible simultaneous measurements 
of the star's $M$ and $R$.  If such measurements are precise, $\eta$
could be deduced to within the accuracy of $\pm20$ MeV,
which would provide a basis for analyzing
more massive neutron stars.

From Fig.\ \ref{fig:etaR}, one can also
observe that the calculated $R$ depends nonlinearly on $\rho_c$ at 
small values of $\eta$, while converging on an approximately linear 
function of $\eta$ at sufficiently large values of $\eta$:
\begin{equation}
  R =  10.32 + 2.57\left(\frac{\eta}{100\, {\rm MeV}}\right)\,{\rm km} . 
  \label{eq:fit-R}
\end{equation}
Note that such nonlinear dependence at small values of $\eta$
arises from the flattened behavior of the corresponding $M$-$R$
relations that can be seen from Fig. \ref{fig:mr}(a), while such
convergence at large values of $\eta$ is related to the vertically
straightened behavior of the corresponding $M$-$R$ relations.

\section{Conclusion}

In this paper, we have succeeded in constructing the theoretical
formulas for the masses, gravitational redshifts, and radii of
low-mass neutron stars as functions of the star's central density
and the new EOS parameter $\eta$ in a manner that is consistent
with empirical masses and radii of stable nuclei.  The value of
$\eta$, which characterizes the stiffness of neutron star matter,
remains unknown, but could be deduced from possible simultaneous
$M$ and $R$ measurements via comparison with our formulas if
the star observed is light enough.  Thus, a firm evidence for
the presence of low-mass neutron stars is first of all desired.
One promising candidate is the neutron star in the high-mass X-ray
binary 4U 1538-52, of which the mass could be significantly low
or even the lowest among stars with known mass if the binary
orbit is eccentric \citep{Rawls2011,PAC2013}. 
The X-ray burster 4U 1724-307 in the
globular cluster Terzan 2 is even more interesting because
the X-ray data from the cooling phase of photospheric radius
expansion bursts apparently allow the object to have a
relatively low mass and still a significantly large radius
\citep{SPRW2011}.  Such conclusions are tentative
partly because of the dependence on the atmosphere models
adopted and partly because of uncertainties in the distance
to the object, but, if valid, might eventually suggest the
$\eta$ value of order or even larger than 130 MeV.

\section*{Acknowledgment}
H.S. is grateful to C. Ishizuka, K. Sumiyoshi, T. Tatsumi, and N. Yasutake for comments. 
H.S., K.I., and A.O. acknowledge the hospitality of the Facility for Rare 
Isotope Beams, where this work was initiated, and thank T. Enoto for
helpful discussions.  This work was supported in part
by Grants-in-Aid for Scientific Research on Innovative Areas through 
No.\ 24105001 and No.\ 24105008 provided by MEXT, in part by Grant-in-Aid for 
Young Scientists (B) through No.\ 24740177 provided by JSPS, in part by the 
Yukawa International Program for Quark-hadron Sciences, and in part by 
Grant-in-Aid for the global COE program ``The Next Generation of Physics,
Spun from Universality and Emergence" from MEXT.



\begin{thebibliography}{99}

\bibitem[Lattimer and Prakash(2004)]{LaPr}
   J. M. Lattimer and M. Prakash, Science {\bf 304}, 536 (2004).

\bibitem[Kiziltan et al.(2013)]{KKYT}
   B. Kiziltan, A. Kottas, M. De Yoreo, and S. E. Thorsett, Astrophys. J. {\bf 778}, 66 (2013).

\bibitem[Guillot et al.(2013)]{GSWR2013}
  S. Guillot, M. Servillat, N. A. Webb, and R. E. Rutledge, Astrophys. J. {\bf 772}, 7 (2013).

\bibitem[Lattimer and Steiner(2013)]{LS2013}
   J. M. Lattimer and A. W. Steiner,
   arXiv:1305.3242

\bibitem[\"{O}zel, Baym, and G\"{u}ver(2010)]{OBG2010}
   F. \"{O}zel, G. Baym, and T. G\"{u}ver,
   Phys. Rev. D {\bf 82}, 101301 (2010).

\bibitem[Steiner, Lattimer, and Brown(2012)]{SLB2012}
  A. W. Steiner, J. M. Lattimer, and E. F. Brown, Astrophys. J. {\bf 765}, L5 (2013).

\bibitem[Tsang et al.(2012)]{exp} 
   M. B. Tsang et al.,
   Phys. Rev. C {\bf 86}, 015803 (2012).

\bibitem[Demorest et al.(2010)]{D2010}
  P. B. Demorest, T. Pennucci, S. M. Ransom, M. S. E. Roberts, and J. W. T. Hessels,
  Nature {\bf 467}, 1081 (2010).

\bibitem[Lattimer and Prakash(2011)]{LP2011}
  J. M. Lattimer and M. Prakash,
  in From Nuclei to Stars: Festschrift in Honor of Gerald E Brown,
  ed. S. Lee (Singapore: World Scientific,  2011), 275

\bibitem[Lattimer(1981)]{L1981}
   J. M. Lattimer, Annu.\ Rev.\ Nucl.\ Part.\ Sci. {\bf 31}, 337 (1981).

\bibitem[Oyamatsu and Iida(2003)]{OI2003}
   K. Oyamatsu and K. Iida,
   Prog. Theor. Phys. {\bf 109}, 631 (2003).

\bibitem{Blaizot1980}
   J. P. Blaizot, Phys.\ Rep. {\bf 64}, 171 (1980).

\bibitem[Roca-Maza et al.(2011)]{Roca-Maza}
   X. Roca-Maza, M. Centelles, X. Vi\~nas, and M. Warda,
   Phys. Rev. Lett. {\bf 106}, 25250 (2011).

\bibitem[Sotani et al.(2012)]{SNIOa} 
   H. Sotani, K. Nakazato, K. Iida, and K. Oyamatsu,
   Phys. Rev. Lett. {\bf 108}, 201101 (2012).
   
\bibitem[Sotani et al.(2013)]{SNIOb} 
   H. Sotani, K. Nakazato, K. Iida, and K. Oyamatsu,
   Mon. Not. R. Astro. Soc. 428, L21 (2013).

\bibitem[Blatt and Weisskopf(1952)]{BW}
   J. M. Blatt and V. F. Weisskopf,
  Theoretical Nuclear Physics (New York: Wiley, 1952).

\bibitem[Oyamatsu (1993)]{O1993}
   K. Oyamatsu, Nucl.\ Phys.\ A {\bf 561}, 431 (1993).

\bibitem[Friedman and Pandharipande(1981)]{FP}
   B. Friedman and V. R. Pandharipande,
   Nucl. Phys. A {\bf 361}, 502 (1981).

\bibitem[Lagaris and Pandharipande(1981)]{LP}
   I. E. Lagaris and V. R. Pandharipande,
   Nucl. Phys. A {\bf 369}, 470 (1981).
   
\bibitem[Oyamatsu and Iida(2007)]{OI2007}
   K. Oyamatsu and K. Iida,
   Phys. Rev. C {\bf 75}, 015801 (2007).

\bibitem{KM2013}
   E. Khan, J. Margueron, Phys. Rev. C {\bf 88}, 034319 (2013).
 


\bibitem[Shen et al.(1998)]{ShenEOSa}
   H. Shen, H. Toki, K. Oyamatsu, and K. Sumiyoshi,
   Nucl. Phys. A {\bf 637}, 435 (1998).

\bibitem[Miyatsu, Yamamuro, and Nakazato(2013)]{Miyatsu}
   T. Miyatsu, S. Yamamuro, and K. Nakazato, Astrophys. J. {\bf 777}, 4 (2013).

\bibitem[Lorenz, Ravenhall, and Pethick(1993)]{FPS}
   C. P.Lorenz, D. G. Ravenhall, and C. J. Pethick,
   Phys. Rev. Lett. {\bf 70}, 379 (1993).

\bibitem[Douchin and Haensel(2001)]{SLy4}
   F. Douchin and P. Haensel,
   Astron. Astrophys. {\bf 380}, 151 (2001).

\bibitem[Goriely, Chamel, and Pearson(2010)]{BSk}
   S. Goriely, N. Chamel, and J. M. Pearson,  
   Phys. Rev. C {\bf 82}, 035804 (2010).

\bibitem[Pearson, Goriely, and Chamel(2011)]{PGC2011}
   J. M. Pearson, S. Goriely, and N. Chamel,
   Phys. Rev. C {\bf 83}, 065810 (2011).

\bibitem[Pearson et al.(2012)]{PCGD2012}
   J. M. Pearson, N. Chamel, S. Goriely, and C. Ducoin,
   Phys. Rev. C {\bf 85}, 065803 (2012).

\bibitem[Chabanat et al.(1997)]{Chabanat1997}
   E. Chabanat, P. Bonche, P. Haensel, J. Meyer, and R. Schaeffer,
   Nucl.\ Phys.\ A {\bf 627}, 710 (1997).

\bibitem[Wiringa, Fiks, and Fabrocini(1998)]{WFF}
   R. B. Wiringa, V. Fiks, and A. Fabrocini,
   Phys.\ Rev.\ C {\bf 38}, 1010 (1988).

\bibitem[Akmal, Pandharipande, and Ravenhall(1998)]{Akmal}
   A. Akmal, V. R. Pandharipande, and D. G. Ravenhall,
   Phys. Rev. C {\bf 58}, 1804 (1998).

\bibitem[Li and Schulze(2008)]{LS}
   Z. H. Li an H. J. Schulze,
   Phys. Rev. C {\bf 78}, 028801 (2008).

\bibitem[Haensel and Potekhin(2004)]{HP2004}
   P. Haensel, and A. Y. Potekhin,
   Astron. Astrphys. {\bf 428}, 191 (2004).

\bibitem[Potekhin et al.(2013)]{PFCPG}
   A. Y. Potekhin, A. F. Fantina, N. Chamel, J. M. Pearson, and S. Goriely,
   Astron. Astrophys. in press (arXiv:1310.0049)    

\bibitem[Gandolfi, Carlson, and Reddy(2012)]{GCR2012}
  S. Gandolfi, J. Carlson, and S. Reddy,
  Phys. Rev. C {\bf 85}, 032801 (2012).

\bibitem[Rutledge et al.(2002)]{RBBPZ2002}
  R. E. Rutledge, L. Bildsten, E. F. Brown, G. G. Pavlov, and V. E. Zavlin,
  Astrophys. J. {\bf 578}, 405 (2002).

\bibitem[Petrov, Antokhina, and Cherepashchuk(2013)]{PAC2013}
   V. S. Petrov, E. A. Antokhina, and A. M. Cherepashchuk,
   Astron. Rep. {\bf 57}, 669 (2013).
    
\bibitem[Rawls et al.(2011)]{Rawls2011}
   M. L. Rawls et al.,
   Astrophys. J. {\bf 730}, 25 (2011).

\bibitem[Suleimanov et al.(2011)]{SPRW2011}
  V. Suleimanov, J. Poutanen, M. Revnivtsev, and K. Werner,
  Astrophys. J.  {\bf 742}, 122 (2011).











   




\end{thebibliography}
%

\vfill\pagebreak


\end{document}